\documentclass[superscriptaddress,amsmath,amssymb,floatfix,onecolumn,notitlepage,aps]{revtex4-1}

\usepackage{amsmath}
\usepackage{amssymb}
\usepackage{amsthm}
\usepackage{mathrsfs}
\usepackage{bm}
\usepackage{graphicx}
\usepackage{caption}
\usepackage{subcaption}
\usepackage{xcolor}
\usepackage[]{hyperref}
\hypersetup{
colorlinks=true,
citecolor=blue,
filecolor=red,
linkcolor=cyan,
urlcolor=cyan
}
\usepackage[section]{placeins}

\numberwithin{equation}{section}

\graphicspath{{./Figures/}}

\theoremstyle{definition}

\newcommand{\R}{\mathbb{R}}

\newcommand{\script}{\mathscr}
\newcommand{\dd}{\mathrm{d}}
\newcommand{\intnice}{\int\!}

\newcommand{\hqcd}{HRG+pQCD}
\newcommand{\lsqrtsigma}{\Lambda_{\overline{\text{MS}}} / \sqrt{\sigma}}

\begin{document}

\title{The equation of state in two-, three-, and four-color QCD at non-zero temperature and density}
\author{Tyler Gorda and Paul Romatschke}
\affiliation{University of Colorado Boulder, Boulder, CO}
\date{\today}

\begin{abstract}
We calculate the equation of state at non-zero temperature and density from first principles in \linebreak 
two-, three- and four-color QCD with two fermion 
flavors in the fundamental and two-index, antisymmetric representation. By matching low-energy results (from a `hadron resonance gas') 
to high-energy results from (resummed) perturbative QCD, we obtain results for the pressure and trace anomaly that are in quantitative 
agreement with full lattice-QCD studies for three colors at zero chemical potential. Our results for non-zero chemical potential at 
zero temperature constitute predictions for the equation of state in QCD-like theories that can be tested by traditional lattice 
studies for two-color QCD with two fundamental fermions and four-color QCD with two two-index, antisymmetric fermions. We find 
that the speed of sound squared at zero temperature can exceed one third, which may be relevant for the phenomenology of high-mass 
neutron stars.
\end{abstract}

\maketitle

\section{Introduction}
\label{section: method}

Knowledge about condensed-matter properties of quantum chromodynamics (QCD) in thermodynamic equilibrium is required for the
interpretation of 
experimental and observational data in cosmology, high-energy nuclear physics and the physics of neutron stars. While tremendous 
progress has been made for the case of high temperature and small baryon densities using direct simulations in lattice QCD 
\cite{Borsanyi:2010cj,Borsanyi:2013bia,Bazavov:2014pvz}, much less is known for the case of small temperature and large densities. 
The reason for this shortcoming is that the so-called sign problem prohibits the direct simulation 
of QCD (which is an SU$(3)$ gauge theory with $n_f=2+1$ fundamental fermionic degrees of freedom) at large density using established 
importance sampling techniques.  While established techniques fail, several recent techniques have been studied that at least 
in principle could permit one to calculate  thermodynamic properties from first principles in QCD at large density. These 
techniques include 
Lefschetz thimbles \cite{Cristoforetti:2012su}, complex Langevin \cite{Aarts:2013uxa,Sexty:2013ica,Aarts:2014bwa,Aarts:2014fsa}, 
strong coupling expansion \cite{deForcrand:2014tha}, and hadron resonance gas plus perturbative QCD (``\hqcd'' in the following) 
\cite{Laine:2006cp,Huovinen:2009yb,Kurkela:2009gj}. 
In this work, we propose a series of `control studies' 
in QCD-like theories (in particular two-color QCD with two fundamental flavors and four-color QCD with two flavors in the two-index, 
antisymmetric representation), which---despite not corresponding to the actual theory of strong interactions realized in nature---have 
the advantage 
of not suffering from a sign problem, and are thus amenable to direct simulations using established lattice-QCD techniques. We then 
proceed to calculate thermodynamic properties in these QCD-like theories in one of the above non-traditional approaches (\hqcd, 
Ref.~\cite{Laine:2006cp,Huovinen:2009yb,Kurkela:2009gj}), which effectively makes predictions for possible future lattice-QCD 
studies that can be used to validate or falsify this \hqcd\ approach.  Since two- and four-color QCD are qualitatively similar to 
three-color QCD, we furthermore expect the level of agreement between lattice 
QCD and \hqcd\ in the two- or four-color cases to be roughly comparable to the three-color QCD case, thus offering an indirect 
validation of non-traditional methods for QCD at large densities.

The paper is organized as follows. 
In Section \ref{sec: pQCD} we review the \hqcd\ method and give the equation of state in pQCD by stating the pressure $P$ as a function 
of temperature $T$ at baryon chemical potential $\mu = 0$ and $P$ as a function of $\mu$ at $T = 0$.  This section is essentially 
a compilation of what has been derived in the
literature.  In Section \ref{sec: HRG} we compute the HRG pressure as a function of $T$ or $\mu$.  The former case is simple and is
derived quickly, whereas the latter is derived in more detail, especially for the theories where the baryons of the theory are bosons.
This section also contains an explanation of how the hadrons in the theories listed above are computed.  Section \ref{sec: matching} 
contains a description of how we perform the matching between these two asymptotic equations of state, and in Section 
\ref{sec: results} we discuss our results.

\section{pQCD equation of state}
\label{sec: pQCD}

In this work, we are interested in calculating the pressure $P$ 
along the $\mu = 0$ and $T = 0$ axes in the theories $(N , n_{f}) = (2 , 2) , 
(3 , 3) , \text{ and } (4 , 2)$ with quarks in the fundamental representation (fundamental) and $(4 , 2)$ with quarks in the two-index, 
antisymmetric representation (antisymmetric).  In order to constrain the pressure of these
theories, we derive the asymptotic behavior for both low and high $T$ or $\mu$ and then 
match these behaviors using basic thermodynamics.  At high $T$ or high $\mu$, the equation of state can be calculated using (resummed) 
pQCD and at low $T$ or $\mu$, the equation of state of the theory is to good approximation \cite{Borsanyi:2010cj,Borsanyi:2013bia,
Bazavov:2014pvz,Langelage:2008dj} that of a HRG (a noninteracting collection of the hadrons of that theory).
In the intermediate regime, the equations of state can be constructed by matching the high/low energy asymptotic behavior using the 
criterion that the pressure $P$ must increase as a function of $T$ or as a function of $\mu$ (see \cite{Vuorinen:2003fs}). More details 
of the matching procedure will be discussed below. Throughout this paper, Boltzmann's constant $k$, Planck's reduced constant $\hbar$, 
and the speed of light $c$ will be set equal to one.

The high-$T$, pQCD equation of state can be calculated by following the equations and procedure of Kajantie \emph{et al.} 
\cite{Kajantie:2002wa,Kajantie:2003ax} and Vuorinen \cite{Vuorinen:2003fs} 
with the resummation modifications described by Blaizot \emph{et al.} \cite{Blaizot:2003iq} (cf.\ Ref.~\cite{Haque:2014rua} for a 
different approach to the resummed pQCD equation of state).  We first define the following group-theory terms to be used in 
all future pQCD expressions:
\begin{align}
	C_{A} &= N, \\
	d_{A} &= N^{2} - 1,
\end{align}
and
\begin{alignat}{2}
	C_{\text{fundamental}} &= \frac{N^{2} - 1}{2 N}, \qquad \qquad &C_{\text{antisymmetric}} &= \frac{(N - 2)(N + 1)}{N}, \\
	T_{\text{fundamental}} &= \frac{n_{f}}{2}, &T_{\text{antisymmetric}} &= \frac{(N - 2)}{2} n_{f} , \\
	d_{\text{fundamental}} &=  N n_{f}, &d_{\text{antisymmetric}} &= \frac{N (N - 1)}{2} n_{f} .
\end{alignat}
In all of the expressions that follow, we let group-theory terms with a subscript $R$ denote the fermionic group-theory 
invariants, which must be replaced by the corresponding fundamental or antisymmetric representation group-theory invariants 
above as needed.  In terms of these group-theory terms, the pQCD pressure at $\mu = 0$ in these theories can be written
\begin{equation}
	P_{\text{pQCD}}(T) = P_{\text{sb}}(T) + P_{\text{hard}}(T) + P_{\text{EQCD}}(T).
\label{eq: pQCD T}
\end{equation}
Here, the $P_{\text{sb}}$ the Stefan-Boltzmann pressure given by
\begin{equation}
	P_{\text{sb}}(T) = \frac{\pi^{2} T^{4}}{45} \left( d_{A} + \frac{7}{4} d_{R} \right).
\label{eq: pQCD sb}
\end{equation}
To 3-loop order, $P_{\text{hard}}$ is given by Braaten and Nieto \cite{Braaten:1995jr} as
\begin{align}
	P_{\text{hard}}(T) &= \frac{\pi^{2} d_{A}}{9} T^{4} \bigg\{ 
			- \left(C_{A} + \frac{5}{2} T_{R} \right) \frac{\alpha_{s}}{4 \pi} \nonumber \\
		& + \bigg( C_{A}^{2} \left[ 48 \ln \frac{\Lambda_{E}}{4 \pi T} - \frac{22}{3} \ln \frac{\overline{\Lambda}}{4 \pi T} 
			+ \frac{116}{5} + 4 \gamma + \frac{148}{3} \frac{\zeta'(-1)}{\zeta(-1)} 
			- \frac{38}{3} \frac{\zeta'(-3)}{\zeta(-3)} \right] 
				\nonumber \\
		& + C_{A} T_{R} \left[ 48 \ln \frac{\Lambda_{E}}{4 \pi T} - \frac{47}{3} \ln \frac{\overline{\Lambda}}{4 \pi T} 
			+ \frac{401}{60} -\frac{37}{5} \ln 2 + 8 \gamma + \frac{74}{3} \frac{\zeta'(-1)}{\zeta(-1)} 
			- \frac{1}{3} \frac{\zeta'(-3)}{\zeta(-3)} \right] \nonumber \\
		& + T_{R}^{2} \left[ \frac{20}{3} \ln \frac{\overline{\Lambda}}{4 \pi T} + \frac{1}{3}
			-\frac{88}{5} \ln 2 + 4 \gamma + \frac{16}{3} \frac{\zeta'(-1)}{\zeta(-1)} - 
			\frac{8}{3} \frac{\zeta'(-3)}{\zeta(-3)} \right] \nonumber \\
		& + C_{R} T_{R} \left[ \frac{105}{4} - 24 \ln 2 \right]\bigg) \left( \frac{\alpha_{s}}{4 \pi} \right)^{2} \bigg\},
\label{eq: pQCD hard}
\end{align}
where $\Lambda_{E}$ is the factorization scale between the hard and soft modes, and $\alpha_{s}$ is the strong coupling constant 
squared over $4 \pi$ in the $\overline{\text{MS}}$ renormalization scheme at 
the scale $\overline{\Lambda} = \sqrt{(2 \pi T)^{2} + (\mu)^{2}}$.  This is given 
by \cite{vanRitbergen:1997va,Kurkela:2009gj}
\begin{equation}
	\alpha_{s}(\overline{\Lambda}) = \frac{4 \pi}{\beta_{0} L} \left(1 - \frac{\beta_{1}}{\beta_{0}^{2}} \frac{\ln L}{L} \right),
		\qquad L = \ln \left( \overline{\Lambda}^{2} / \Lambda^{2}_{\overline{\text{MS}}} \right),
\end{equation}
with 
\begin{equation}
	\beta_{0} = \frac{11}{3} C_{A} - \frac{4}{3} T_{R}, \quad \beta_{1} = \frac{34}{3} C_{A}^{2} - 4 C_{R} T_{R} 
			- \frac{20}{3} C_{A} T_{R},
\end{equation}
where $\Lambda_{\overline{\text{MS}}}$ is the $\overline{\text{MS}}$ renormalization point (to be set later).  In all our results, 
we set $\Lambda_{E} = \overline{\Lambda}$ and vary $\overline{\Lambda}$ about the aforementioned value by a factor of two 
(cf.\ the end of Section \ref{sec: matching}). Finally, 
$P_{\text{EQCD}}$ is given by
\begin{align}
	P_{\text{EQCD}}(T) = \frac{d_{A}}{4 \pi} T \bigg( \frac{1}{3} m_{E}^{3} - \frac{C_{A}}{4 \pi} \left( \ln \frac{\Lambda_{E}}{2 m_{E}} + \frac{3}{4} \right) 
		g_{E}^{2} m_{E}^{2} - \left( \frac{C_{A}}{4 \pi} \right)^2 \left( \frac{89}{24} - \frac{11}{6} \ln 2 + \frac{1}{6} \pi^{2} \right) 
		g_{E}^{4} m_{E} \bigg),
\label{eq: pQCD EQCD}
\end{align}
where
\begin{align}
	m_{E}^{2} &= \frac{4 \pi}{3} \alpha_{s} T^{2} \bigg\{ C_{A} + T_{R} \nonumber \\ 
		&+ \Big[ C_{A}^{2} \left( \frac{5}{3} + \frac{22}{3} \gamma + \frac{22}{3} \ln \frac{\overline{\Lambda}}{4 \pi T} 
			\right) + C_{A} T_{R} \left( 3 - \frac{16}{3} \ln 2 + \frac{14}{3} \gamma + \frac{14}{3} \ln 
			\frac{\overline{\Lambda}}{4 \pi T} \right) 
				\nonumber \\
		&+ T_{R}^{2} \left( \frac{4}{3} - \frac{16}{3} \ln 2 - \frac{8}{3} \gamma 
			- \frac{8}{3} \ln \frac{\overline{\Lambda}}{4 \pi T} \right) 
			- 6 C_{R} T_{R} \Big] \left( \frac{\alpha_{s}}{4 \pi} \right) \bigg\},
\end{align}
and
\begin{equation}
	g_{E}^2 = 4 \pi \alpha_{s} T.
\end{equation}

The $T = 0$, pQCD equation of state is more straightforward in the sense that resummation of the strict perturbative series is not 
required. The result is given in Ref.~\cite{Vuorinen:2003fs} by
\begin{align}
	P_{\text{pQCD}}(\mu) & = \frac{1}{4 \pi^{2}} \Bigg( \sum_{f} \mu_{f}^{4} \bigg\{ \frac{d_{R}}{3 n_{f}} - d_{A} 
			\left( \frac{2 T_{R}}{n_{f}} \right)
			\left( 
			\frac{\alpha_{s}}{4 \pi} \right) - d_{A} \left( \frac{2 T_{R}}{n_{f}} \right)
			\left( \frac{\alpha_{s}}{4 \pi} \right)^{2} \bigg[ 
			\frac{2}{3}(11 C_{A} - 4 T_{R}) \ln \frac{\overline{\Lambda}} {\mu_{f}} + \frac{16}{3} \ln 2 \nonumber \\
		& + \frac{17}{4} \left( \frac{C_{A}}{2} - C_{R} \right) + \frac{1}{36}(415 - 264 \ln 2) C_{A} - \frac{8}{3} \left( 
			\frac{11}{6} - \ln 2 \right) T_{R} \bigg] \bigg\} \nonumber \\
		& - d_{A} \left( \frac{2 T_{R}}{n_{f}} \right)\left( \frac{\alpha_{s}}{4 \pi} \right)^{2} \left\{ \left( 2 \ln 
			\frac{\alpha_{s}}{4 \pi} - \frac{22}{3} + 
			\frac{16}{3} \ln 2 \, (1 - \ln 2) + \delta + \frac{2 \pi^{2}}{3} \right) (\bm{\mu}^{2})^{2} 
			+ F(\bm{\mu}) \right\} \Bigg) \nonumber \\
		& + \mathcal{O} (\alpha_{s}^{3} \ln \alpha_{s}),
\label{eq: pQCD mu}
\end{align}
where the sum is over all the quark flavors in the theory, $\mu_{f}$ is the $f$-quark chemical potential, $\bm{\mu}^{2} = \sum_{f} 
\mu_{f}^{2}$, and 
\begin{align}
	F(\bm{\mu}) & = -2 \bm{\mu}^{2} \left( \frac{2 T_{R}}{n_{f}} \right)\sum_{f} \mu_{f}^{2} \ln \frac{\mu_{f}^{2}}{\bm{\mu}^{2}} 
			+ \frac{2}{3} \left( \frac{2 T_{R}}{n_{f}} \right)^{2} \sum_{f > g} 
			\bigg\{ (\mu_{f} - \mu_{g})^{2} \ln \frac{|\mu_{f}^{2} - \mu_{g}^{2}|}{\mu_{f} \mu_{g}} \nonumber \\
		& + 4 \mu_{f} \mu_{g} (\mu_{f}^{2} + \mu_{g}^{2}) \ln \frac{(\mu_{f} + \mu_{g})^{2}}{\mu_{f} \mu_{g}} - 
			(\mu_{f}^{4} - \mu_{g}^{4}) \ln \frac{\mu_{f}}{\mu_{g}} \bigg\},
\label{eqn: F-function}
\end{align}
with the constant $\delta$ having the value $\delta = −0.85638320933$.  In what follows we always set 
all of the quark chemical potentials equal to each other, so that $\mu_{f} = \mu / N_{b}$ for each flavor $f$, where $N_{b}$ is the
number of quarks in a baryon.  Note that this means that some of the terms in \eqref{eqn: F-function} do not contribute.

\section{Hadron resonance gas equation of state}
\label{sec: HRG}

The low-$T$ pressure in these theories is given by considering the system to be a free gas of hadrons.  Moreover, the statistics of 
the hadrons may be ignored, so that the distribution functions may all be assumed to be Boltzmann factors.  In that case, we have
\begin{equation}
	P_{\text{HRG}}(T) = T \sum_{i \in H} {g_{i}} \intnice \frac{\dd^{3}p}{(2 \pi)^{3}} e^{- \sqrt{\mathbf{p}^{2} + m_{i}^{2}} / T}
		= T^{4} \sum_{i \in H} \frac{g_{i}}{2 \pi^{2}} \left( \frac{m_{i}}{T} \right)^{2} K_{2} \! \left( \frac{m_{i}}{T} \right),
\label{eq: HRG}
\end{equation}
where here the sum is over the hadron spectrum of the theory; $g_{i}$ and $m_{i}$ are the degeneracy and the mass of the $i$th
particle, respectively; $\mathbf{p} = | \vec{p} \,|$; and $K_{2}$ is a modified Bessel function of the second kind.

The low-$\mu$ pressure can be calculated in a similar way, but in this case the statistics of the particles cannot be ignored.  For 
$T = 0$ and $\mu > 0$, the only particles that contribute to the partition function are particles containing no antiquarks,
which we denote by $B$. In all
the fundamental theories, these are simply the baryons, whereas for the antisymmetric theory there are more particles fitting this 
description (see
below).  As such, in this section we shall refer to all the particles in $B$ as baryons.  
Taking the $T \to 0$ limit in the fermionic-baryon ($\eta = 1$) or bosonic-baryon ($\eta = -1$) case yields
\begin{align}
	P_{\text{HRG}}(\mu) & = \lim_{T \to 0} T \sum_{i \in B} g_{i} \eta \intnice \frac{\dd^{3}p}{(2 \pi)^{3}} \ln \left(1 + \eta \,
			e^{\left(\mu r_{i} - \sqrt{\mathbf{p}^{2} + m_{i}^{2}}\right) / T} \right) \nonumber \\
		& = \sum_{i \in B} g_{i} \eta \intnice \frac{\dd^{3}p}{(2 \pi)^{3}} \ln \left[ \lim_{T \to 0} \left(1 + \eta \,
			e^{\left(\mu r_{i} - \sqrt{\mathbf{p}^{2} + m_{i}^{2}}\right) / T} \right)^{\! T} \right] \nonumber \\
		& =  \sum_{i \in B} g_{i} \eta \int_{0}^{\sqrt{(\mu r_{i})^{2} - m_{i}^{2}}} \frac{\mathbf{p}^{2} \dd \mathbf{p}}
			{2 \pi^{2}} \left(\mu r_{i} - \sqrt{\mathbf{p}^{2} + m_{i}^{2}}\right) \theta( \mu r_{i} - m_{i}) \nonumber \\
		& = \eta \sum_{i \in B} \frac{g_{i}}{48 \pi^{2}} \left[ \mu r_{i} \sqrt{(\mu r_{i})^{2} - m_{i}^{2}} 
			(2 (\mu r_{i})^{2} - 5 m_{i}^{2}) + 3 m_{i}^{4} \cosh^{-1} \left( \frac{m_{i}}{\sqrt{(\mu r_{i})^{2} 
			- m_{i}^{2}}} 
			\right) \right] \theta( \mu r_{i} - m_{i}),
\label{eq: fermion mu}
\end{align}
where here $\theta$ is the Heaviside step function and $r_{i} = N_{i} / N_{b}$, with $N_{i}$ being the number of quarks in the $i$th
particle.  This result is correct for the fermionic-baryon case, but this formula gives
negative $P$ in the bosonic-baryon case when $\mu > \min_{i \in B} m_{i} / r_{i}$.  This is because, in the bosonic case, a condensate 
of the $i$th baryon forms when $\mu = m_{i} / r_{i}$. (This has been numerically investigated in the two-color case
by Hands \emph{et al.} \cite{Hands:2006ve} and analytically by Kogut \emph{et al.} \cite{Kogut:2000ek} in all QCD-like theories 
with pseudoreal fermions.)  In fact, in the completely-noninteracting case it is nonsensical for $\mu$ 
to exceed $\min_{i \in B} m_{i} / r_{i}$.  Since the hadrons in these theories are composite particles, they are not truly 
noninteracting, and we can have $\mu > \min_{i \in B} m_{i} / r_{i}$. 

To make sense of this case, we consider the bosons as a (complex) quantum field $\Phi$ with a $|\Phi|^{4}$ repulsive interaction. 
For simplicity, we consider each baryon to be an independent field, and we examine the case of a scalar field (degeneracies may easily
be incorporated at the end).  A single baryon then has the Lagrangian density (in the mostly minus convention)
\begin{equation}
	\script{L} = (\partial_{\mu} \Phi^{\dagger})(\partial^{\mu} \Phi) - m^{2} \Phi^{\dagger} \Phi - \lambda (\Phi^{\dagger} \Phi) 
		(\Phi^{\dagger} \Phi),
\end{equation}
with $\lambda > 0$.  Following Kapusta and Gale \cite{Kapusta:2006pm}, we introduce a baryon chemical potential $\mu r$ and explicitly
factor out the zero momentum mode
\begin{equation}
	\Phi = \xi + \chi,
\end{equation}
where $\xi \in \R$ is a constant and the constant Fourier component of $\chi$ satisfies $\chi_{n = 0}(\mathbf{p} = 0) = 0$.  One may 
think of $\xi$ as the condensate field and $\chi$ as the fluctuations about the vacuum state.  We also write the fluctuations in terms 
of the normalized real and imaginary parts
\begin{equation}
	\chi = \frac{1}{\sqrt{2}} (\chi_{1} + i \chi_{2}).
\end{equation}
In terms of these new variables, the Euclidean Lagrangian density becomes
\begin{align}
	\script{L} & = -\frac{1}{2} \left( \frac{\partial \chi_{1}}{\partial \tau} - i \mu r \chi_{2} \right)^{2} 
			- \frac{1}{2} \left( \frac{\partial \chi_{2}}{\partial \tau} + i \mu r \chi_{1} \right)^{2} 
			- \frac{1}{2} \nabla^{2} \chi_{1} - \frac{1}{2} \nabla^{2} \chi_{2} \nonumber \\
		& -\frac{1}{2}(6 \lambda \xi^{2} + m^{2}) \chi_{1}^{2} -\frac{1}{2}(2 \lambda \xi^{2} + m^{2}) \chi_{2}^{2} 
			- U(\xi) + \script{L}_{\text{I}},
\label{eq: lagrangian}
\end{align}
where $\tau$ is the Euclidean time, $\script{L}_{\text{I}}$ contains interacting terms in $\chi$ (which we henceforth ignore), and
\begin{equation}
	U(\xi) = (m^{2} - (\mu r)^{2}) \xi^{2} + \lambda \xi^{4}.
\label{eq: vacuum}
\end{equation}
We thus see from \eqref{eq: vacuum} that for $\mu r < m$ the state $\xi_{0} = 0$ is the stable vacuum and \eqref{eq: lagrangian} 
describes a system of particles and antiparticles of equal masses.  However, for $\mu r > m$, the stable vacuum becomes 
\begin{equation}
	\xi_{0}^{2} = \frac{(\mu r)^{2} - m^{2}}{2 \lambda},
\end{equation}
and \eqref{eq: lagrangian} describes a collection of two particles with differing masses: 
$\overline{m}_{1}^{2} = 3 (\mu r)^{2} - 2 m^{2}$ and $\overline{m}_{2}^{2} = (\mu r)^{2}$ respectively. Because of the chemical 
potential, the dispersion relation of the latter is gapless, and Goldstone's theorem is satisfied.  At zero temperature, the 
pressure is simply
\begin{equation}
	P_{\text{HRG}}(\mu) = U(\xi)\big|_{\xi = \xi_{0}} = \frac{1}{4 \lambda} ((\mu r)^{2} - m^{2})^{2} \theta(\mu r - m).
\end{equation}
This gives us the dependence of the pressure on $\mu$, but we still have not set the coupling constant $\lambda$.  We set it as 
follows.  According to \eqref{eq: pQCD mu}, we see that the Fermi--Dirac pressure for a quark in the theory $(N , n_{f})$ with 
representation $R$ becomes
\begin{equation}
	P_{\text{fd}} = \frac{d_{R}}{12 \pi^{2}} \left( \frac{\mu r}{N_{b}} \right)^{4},
\end{equation}
so that for a single degree of freedom (recalling that a fermionic quark has two degrees of freedom) one has
\begin{equation}
	P_{\text{fd}} = \frac{1}{24 \pi^{2} N_{b}^{4}} (\mu r)^{4}.
\end{equation}
Thus, in order for $P_{\text{HRG}} \to P_{\text{fd}}$ when $\mu \to \infty$, we must have for a single scalar baryon
\begin{equation}
	P_{\text{HRG}}(\mu) = \frac{1}{24 \pi^{2} N_{b}^{4}} ((\mu r)^{2} - m^{2})^{2} \theta(\mu r - m),
\end{equation}
and so for a theory with bosonic baryons we have
\begin{equation}
	P_{\text{HRG}}(\mu) = \sum_{i \in B} \frac{g_{i}}{24 \pi^{2} N_{b}^{4}} ((\mu r_{i})^{2} - m_{i}^{2})^{2} 
			\theta(\mu r_{i} - m_{i}).
\label{eq: bosonic mu}
\end{equation}

\subsection{Determining the hadron spectrum}

For the three-color $(N , n_{f}) = (3 , 3)$ fundamental case, we use the real world spectrum of hadrons up to 2.25 GeV 
\cite{Agashe:2014kda}.  For the two- and four-color theories with two fundamental quarks and the four-color theory with two 
antisymmetric quarks, we determine the hadrons using group-theoretic 
arguments and Fermi statistics (in the case of objects composed of quarks only).  We explicitly ignore the glueballs in these theories
because they tend to be more massive than the lightest hadrons \cite{Meyer:2004jc}. For the two- and four-color theories, we set the 
scale using the string tension $\sqrt{\sigma}$ and the relation between the string tension and 
the $\overline{\text{MS}}$ renormalization scale $\Lambda_{\overline{\text{MS}}}$ given in Ref.~\cite{Lucini:2008vi}.  
However, the ratios $\Lambda_{\overline{\text{MS}}} / \sqrt{\sigma}$ 
given in the aforementioned reference are for the pure-gauge theories.  To remedy this, we scale these ratios by 
$\Lambda^{N = 3}_{\overline{\text{MS}}}(n_{f}\!=\!2) / \Lambda^{N = 3}_{\overline{\text{MS}}}(n_{f}\!=\!0)$, determined from 
Ref.~\cite{Fritzsch:2012wq}.  These lead to the values 
\begin{equation}
	\Lambda^{N = 2}_{\overline{\text{MS}}}(n_{f}\!=\!2) / \sqrt{\sigma} = 1.032 \quad \text{and} \quad 
		\Lambda^{N = 4}_{\overline{\text{MS}}}(n_{f}\!=\!2) / \sqrt{\sigma} = 0.723
\end{equation} 
for the fundamental theories. For the three-color theory, we use $\Lambda_{\overline{\text{MS}}} = 0.378 \text{ GeV}$, 
as in \cite{Kurkela:2009gj}.

For the four-color antisymmetric theory, we were unable to locate a result for $\Lambda^{N = 4}_{\overline{\text{MS}}}(n_{f}\!=\!2)
/\sqrt{\sigma}$ 
from the lattice in the literature.  Since some of the group-theory terms for the antisymmetric theory scale more strongly with
the number of colors than the corresponding terms in the fundamental theory, it seems reasonable to expect that 
$\Lambda_{\overline{\text{MS}}}$
will scale differently with the number of quark flavors in the antisymmetric theory than in the fundamental theory.  Moreover, 
it would be most accurate to view $\Lambda_{\overline{\text{MS}}} / \sqrt{\sigma}$ as a free parameter in our \hqcd\ scheme that
must be determined independently from the lattice.  In light of these considerations, we have decided to use both the pure-glue value
\cite{Lucini:2008vi}
\begin{equation}
		\Lambda^{N = 4}_{\overline{\text{MS}}} / \sqrt{\sigma} = 0.527,
\end{equation}
and the previously-given value of $\Lambda^{N = 4}_{\overline{\text{MS}}}(n_{f}\!=\!2) /\sqrt{\sigma}$ that we use for the four-color 
fundamental theory for the four-color antisymmetric theory, with the expectation that the true value will lie somewhere near
this range.

For both the two- and four-color cases, the mesons are taken to be the analogues of the flavorless mesons that exist
in the real world (up to a mass of about 2 GeV) whose masses are written in multiples of the string tension $\sigma_{\text{SU}(3)}
= (420 \text{ MeV})^{2}$.  In the two-color case, we mainly use the analogues of the real-world mesons, substituting the two-color 
masses calculated by Bali \emph{et al.} \cite{Bali:2013kia} when available. (We also note here that the $\mu$-dependence of the 
two-color spectrum has been studied numerically in Ref.~\cite{Hands:2007uc} and analytically in Ref.~\cite{Kogut:2000ek}, though 
we do not need this $\mu$-dependence for our \hqcd\ scheme.) 

We now discuss in some detail how the non-meson objects in these three cases are determined.  For convenience and as a summary of 
these sections, we list tables for all of the particles that we have included in the SU(2) and SU(4) cases in Appendix 
\ref{sec: particle tables}.

\subsubsection{Two-color case}

In two-color QCD, the baryons are composed of two quarks with the added simplicity that the masses are degenerate with the corresponding
mesons made from the same quarks \cite{Lewis:2011zb}.  Thus, the mass spectrum of the baryons is the same as the mass spectrum of 
mesons.  However, there are fewer mesons than baryons, for there is an additional constraint imposed by Fermi statistics in the case 
of the baryons. Since we may view the two massless quarks as part of an isospin doublet, one sees that exchanging the two internal 
quarks in a baryon causes the wavefunction to become multiplied by
\begin{equation}
	(-1)^{1 + L + S + I}.
\end{equation}
In this equation, $L$ is the angular momentum quantum number, $S$ is the spin, and $I$ is the isospin, with the additional $1$ due to the
fact that the quarks are in an antisymmetric color singlet.  We thus see that for even $L$
the spin and isospin must be equal ($S = 0$ implies $I = 0$ and $S = 1$ implies $I = 1$), and for odd $L$ they must be the opposite 
in order
to have a totally antisymmetric wavefunction. (Even though the composite baryon is itself a boson in two-color QCD, it is still a 
multiparticle state of fundamental fermions.) This information is enough to determine the set of hadrons in \eqref{eq: HRG}.

\subsubsection{Four-color fundamental case}

Baryons in four-color QCD with fundamental fermions consist of four quarks.  In this case, to determine the masses $M$ we use the 
large-$N$ expansion
\begin{equation}
	M(J) = N A + \frac{J (J + 1)}{N} B,
\label{eq: large N masses}
\end{equation}
where $J$ is the total angular momentum of the baryon, and $A$, $B$ are constants independent of $N$ 
\cite{Adkins:1983ya,Jenkins:1993zu}.  As pointed out
by DeGrand \cite{DeGrand:2012hd} and demonstrated by Appelquist \emph{et al.} \cite{Appelquist:2014jch}, a term independent 
of $N$ could be used for better agreement. However, we have no way to set the value of that term and thus do not include it.

We find the possible values of $J$ beginning with the ground-state baryons of zero orbital angular momentum.  Since we still have a
isospin doublet of massless, spin-one-half quarks, we only need the group-theory expression
\begin{equation}
	2 \otimes 2 \otimes 2 \otimes 2 = 5_{S} \oplus 3_{M} \oplus 3_{M} \oplus 3_{M} \oplus 1_{A} \oplus 1_{A},
\end{equation}
where the $5_{S}$ state is fully symmetric, the $3_{M}$ states are are symmetric in three of the four quarks and antisymmetric in the 
other, and the $1_{A}$ states are pairwise antisymmetric. Since, again, the quarks are in an antisymmetric color singlet, it must be the 
case that they are in a \emph{symmetric} combination of spin and flavor.  This means that there is a spin-2 quintet, a spin-1 
triplet, and a spin-0 singlet of ground-state baryons.

We may also determine the first excited states in this simple manner by realizing that for this four-body problem there are three 
relevant 
orbital-angular-momentum quantum numbers and the first excited state corresponds to when exactly one of them is one.  In order to
still be in a completely antisymmetric state, either the spin or the flavor state must now be in one of the $3_{M}$ states while the other
must be in a $5_{S}$ state.  This means that there are a quintet of particles with $S = 1$ and a triplet of particles with $S = 2$.  
Combining these with an orbital angular momentum $L = 1$ yields three baryonic quintets with $J = 0, 1, 2$ and three baryonic 
triplets with $J = 1, 2, 3$. We did not determine the baryons for any higher excited states.

\subsubsection{Four-color antisymmetric case} 

The hadron spectrum in the four-color theory with two antisymmetric quarks consists of two-quark objects: mesons and diquarks; 
four-quark objects: tetraquarks, di-mesons, and diquark-mesons; and six-quark baryons 
\cite{DeGrand:2014U,DeGrand:2014cea,Bolognesi:2006ws}.
Since the antisymmetric representation is real, the arguments of Ref.~\cite{Lewis:2011zb} carry through here and one may conclude that 
all two-quark objects with the same quark content have degenerate masses and that the same holds for the four-quark objects.  
In addition, the four-quark objects have a mass equal to the sum of their constituent two-quark-object masses \cite{DeGrand:2014U}.  
Because of this mass degeneracy, 
we need not determine how all of the four-quark-object degrees of freedom break up into spin and isospin multiplets; rather, we may
simply combine the two-quark-object degrees of freedom in every possible way.  One major difference from
the two-color case, however, is that in the four-color theory with antisymmetric quarks the color-singlet state for diquarks 
is \emph{symmetric}.
This means that the spin-isospin locking in this theory is the opposite of the locking in the two-color theory.  That is, for odd $L$ 
the spin and isospin must be equal ($S = 0$ implies $I = 0$ and $S = 1$ implies $I = 1$), and for even $L$ they must be opposite.
As for the six-quark baryons, we again use the large-$N$ expression
\eqref{eq: large N masses}, but with $N$ replaced by $N_{b} = 6$, the number of quarks in the baryon.  We include only the 
ground-state baryons, where isospin and spin are locked as $I = J =$ 3, 2, 1, and 0 \cite{DeGrand:2014U}.

\subsection{Chiral symmetry breaking and the Nambu--Goldstone bosons}
\label{sec: ChSB}

The lowest-mass particles in all of the aforementioned theories are precisely zero at zero quark mass.  This can be understood in 
terms of
the pattern of chiral symmetry breaking in these theories \cite{Peskin:1980gc}.  Consider an SU($N$) gauge theory with 
$n_{f}$ massless fermions.  For fermions in a complex representation (such as in the cases $N \ge 3$ with fundamental fermions), 
the Lagrangian density possesses the symmetry $\text{U}(n_{f}) \otimes \text{U}(n_{f})$,
corresponding to the separate left- and right-handed chiral symmetries, and for real representations (such as any $N$ with 
adjoint fermions or $N = 4$ with antisymmetric fermions) or pseudoreal representations (such as in $N = 2$ with fundamental 
fermions), the Lagrangian density possesses the 
larger symmetry $\text{U}(2 n_{f})$.  In all of these cases, the axial U(1) symmetry is broken by an anomaly, and 
the remaining symmetries are spontaneously broken in the following ways. For fermions in a complex representation:
\begin{equation}
	\text{SU}(n_{f}) \otimes \text{SU}(n_{f}) \to \text{SU}(n_{f});
\end{equation}
for fermions in a real representation:
\begin{equation}
	\text{SU}(2 n_{f}) \to \text{O}(2 n_{f});
\end{equation}
and for fermions in a pseudoreal representation:
\begin{equation}
	\text{SU}(2 n_{f}) \to \text{Sp}(2 n_{f}).
\end{equation}
(See Refs.~\cite{Peskin:1980gc,Kogut:2000ek} for more details.)  The generators of the broken symmetries become massless 
Nambu--Goldstone bosons.  Since SU($n_{f}$) has $n_{f}^{2} - 1$ generators,  O($n_{f}$) has $n_{f} (n_{f} - 1) / 2$ generators, 
and Sp($n_{f}$) has $n_{f} (n_{f} + 1) / 2$ generators, we see that in the three-color, three fundamental-quark case there 
will be 8 Nambu--Goldstone bosons 
(a meson octet); in the four-color, two antisymmetric-quark case there will be 9 Nambu--Goldstone bosons (a triplet each of mesons, 
diquarks, and antidiquarks); and in the two-color, two fundamental-quark case there will be 5 Nambu--Goldstone bosons (a triplet of 
mesons, a diquark, and an antidiquark).

In addition, recall that if the quarks in these theories are not precisely massless, then the massless Nambu--Goldstone bosons will become 
instead small-mass, pseudo-Nambu--Goldstone bosons.  In our spectra, we are free to vary the mass of these lightest particles to see what 
effects this will have on the equation of state of the theories.  This is especially interesting for lattice practitioners.  We discuss
this further in Section \ref{sec: results}.

\section{Matching the pQCD and HRG equations of state}
\label{sec: matching}

To match the two asymptotic equations of state, we employ the same technique on the $T$ axis as on the $\mu$ axis.  As such, let us 
introduce the symbol $F$ to stand for either $T$ or $\mu$ so that we may discuss the matching in full generality.  

To perform the matching, we assume that at the phase-transition point the pressures of the two phases are equal; 
and we use the thermodynamic constraints that the pressure of a system must increase with $F$
\begin{equation}
	P(F + \Delta F) \ge P(F),
\label{eq: pressure increase}
\end{equation}
and that above a phase-transition point, the physical phase is the one with the higher pressure.  We also add a bag constant $B$ to the pQCD 
pressure so that
\begin{equation}
	P_{\text{pQCD}}(F) = P_{\text{pQCD}}^{0}(F) + B,
\end{equation}
where $P_{\text{pQCD}}^{0}$ is given by either \eqref{eq: pQCD T}-\eqref{eq: pQCD EQCD} or \eqref{eq: pQCD mu}.  In the plots that follow, 
we solve the following set of two equations with two unknowns (for a given $\overline{\Lambda}$):
\begin{align}
	P_{\text{HRG}}(F_{0}) &= P_{\text{pQCD}}(F_{0} , B_{0}),
\label{eq: matching A} \\
	\frac{\dd P_{\text{HRG}}(F)}{\dd F} \bigg|_{F = F_{0}} &= \frac{\partial P_{\text{pQCD}} (F , B_{0})}{\partial F} 
		\bigg|_{F = F_{0}}.
\label{eq: matching B}
\end{align}
The second of these equations amounts to assuming that the phase transition is of second order. Less restrictive schemes can be implemented
as well, such as truncating the HRG and pQCD equations of state far from the transition region and interpolating using the thermodynamic
constraint \eqref{eq: pressure increase} as well as other phenomenological assumptions (see Refs.~\cite{Kurkela:2014vha,Kojo:2014rca} for examples of this). By 
varying $\overline{\Lambda}$ between $\pi T$ and $4 \pi T$ for the case $F = T$ and between $\mu / 2$ and $2 \mu$ in the case $F = \mu$,
\eqref{eq: matching A}-\eqref{eq: matching B} allows us to obtain a region of possible equations of state in the $(F , P)$ plane 
for each theory.

\section{Results}
\label{sec: results}

\begin{figure}[h]
\centering
\includegraphics[width = 0.49\linewidth]{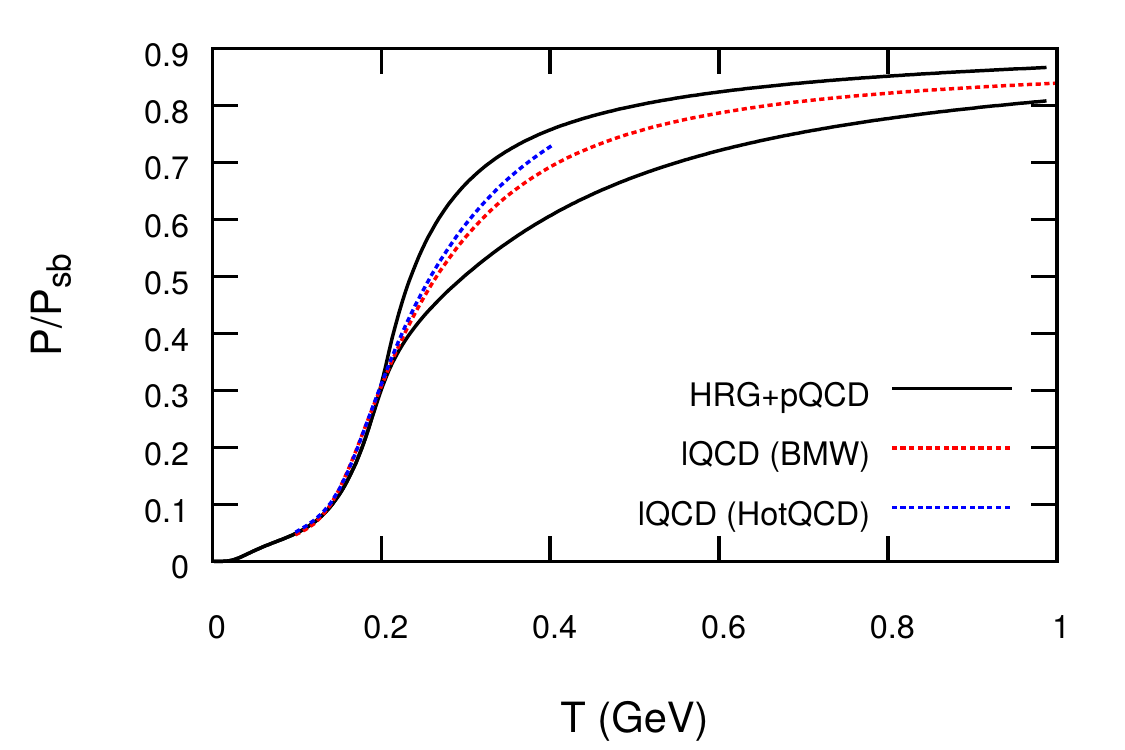}
\includegraphics[width = 0.49\linewidth]{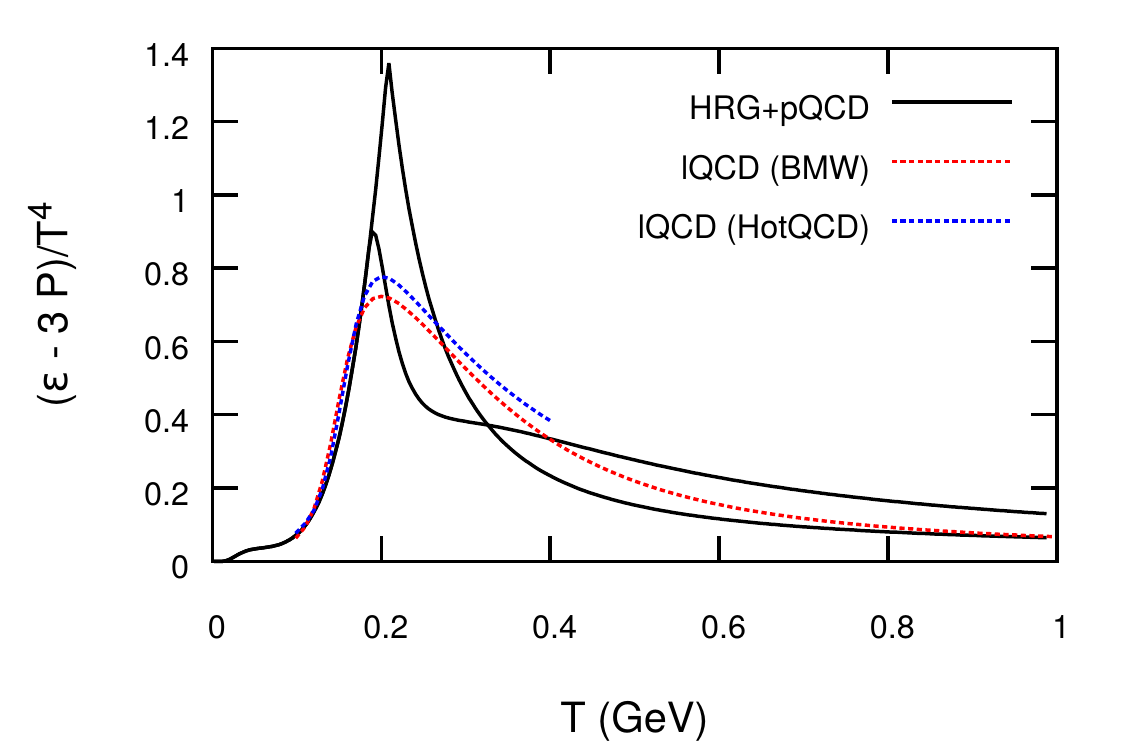}
\caption{ Normalized pressure (left) and trace anomaly (right) at $\mu = 0$ for the 
 three-color, three-massless-quark case from \hqcd\ in comparison to lattice-QCD data from the Budapest--Marseille--Wuppertal 
 Collaboration 
 \cite{Borsanyi:2013bia} and the HotQCD Collaboration \cite{Bazavov:2014pvz}.}
\label{fig: Three Color Agree}
\end{figure}

In Figure~\ref{fig: Three Color Agree}, we overlay the bands for the pressure and trace anomaly $\epsilon-3 P$ (with $\epsilon$ 
the energy density) at $\mu = 0$ that we calculate in the 
three-color, three-massless-quark case with lattice data from the Budapest--Marseille--Wuppertal Collaboration \cite{Borsanyi:2013bia} 
and the HotQCD Collaboration \cite{Bazavov:2014pvz} in their respective regions of validity.  We observe  that 
the lattice data agree reasonably well with the band resulting from our \hqcd\ calculation, both for the pressure as well as for the 
trace anomaly. 

\begin{figure}[h]
\centering
\includegraphics[width = 0.49\linewidth]{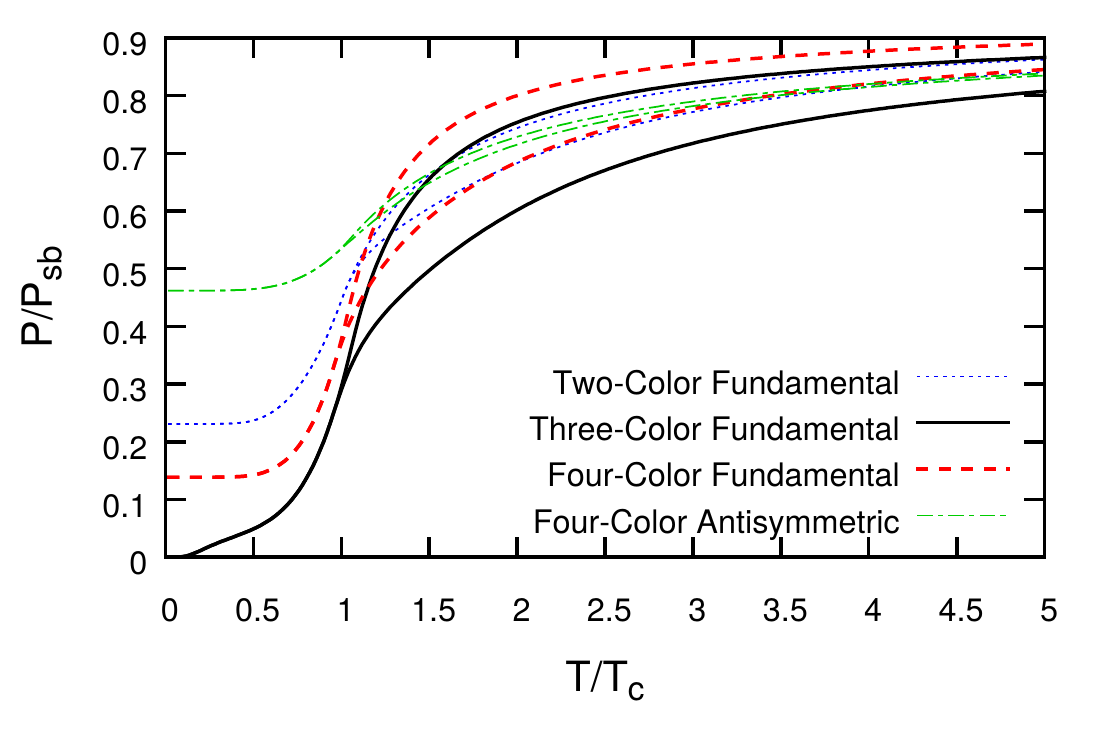}
\includegraphics[width = 0.49\linewidth]{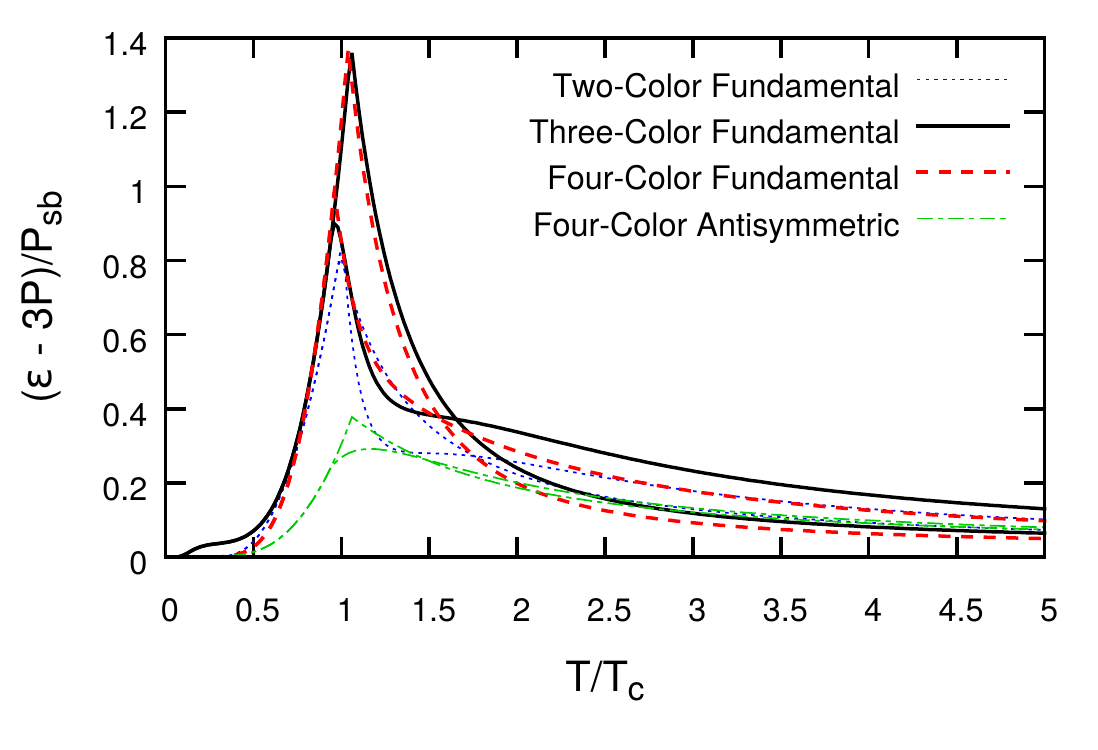}
\caption{Normalized pressure (left) and trace anomaly (right) at $\mu = 0$ for the 
 two-color, three-color, four-color fundamental, and four-color antisymmetric theories in \hqcd.  Note that the $T$ axis has been scaled by 
 the critical temperature (see main text).}
\label{fig: Overlay EoS}
\end{figure}

In Figure~\ref{fig: Overlay EoS}, we  show the \hqcd\ pressure and trace-anomaly bands at $\mu = 0$ 
for all four theories with the $T$ axis scaled by the critical temperature $T_{c}$, which we define to be the 
average of the matching temperatures for the upper and lower edge of our pQCD band to the HRG equation of state. $T_c$ should 
thus be regarded as an estimate of the confinement-deconfinement critical temperature. We list the explicit 
values obtained in \hqcd\ in Table \ref{tab: Tc MuC}. We see that once the temperature axis has been scaled by $T_c$, 
all the theories show similar behavior both for the pressure and trace anomaly, a phenomenon that is well-known from pure-gauge 
theories \cite{Panero:2009tv}.  The differences at low temperatures are due to the different numbers of Nambu--Goldstone bosons
with zero quark mass in the two-color and four-color theories (see Section~\ref{sec: ChSB} or Appendix~\ref{sec: particle tables})
and the fact that in the real world there are only pseudo-Nambu--Goldstone bosons. We tested this by increasing the mass of the 
lightest (now pseudo-) Nambu--Goldstone bosons, which qualitatively changed the shape of the pressure curves until they matched that
of the real-world, three-color theory.

\begin{figure}[h]
\centering
\includegraphics[width = 0.49\linewidth]{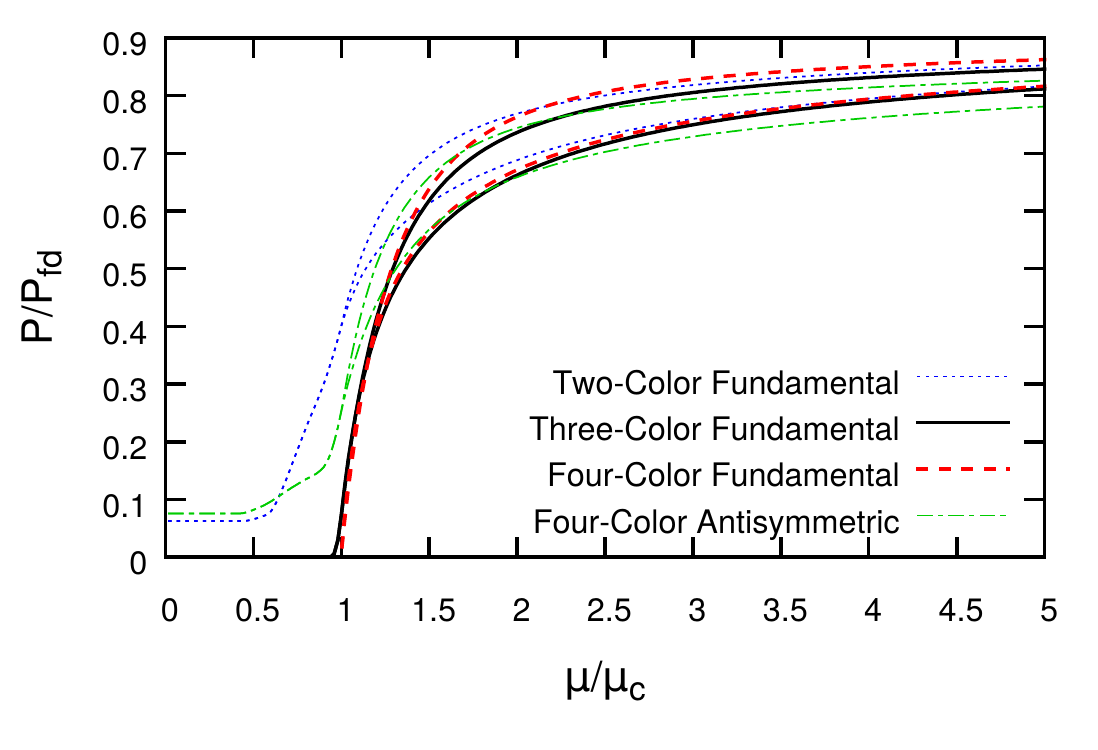}
\includegraphics[width = 0.49\linewidth]{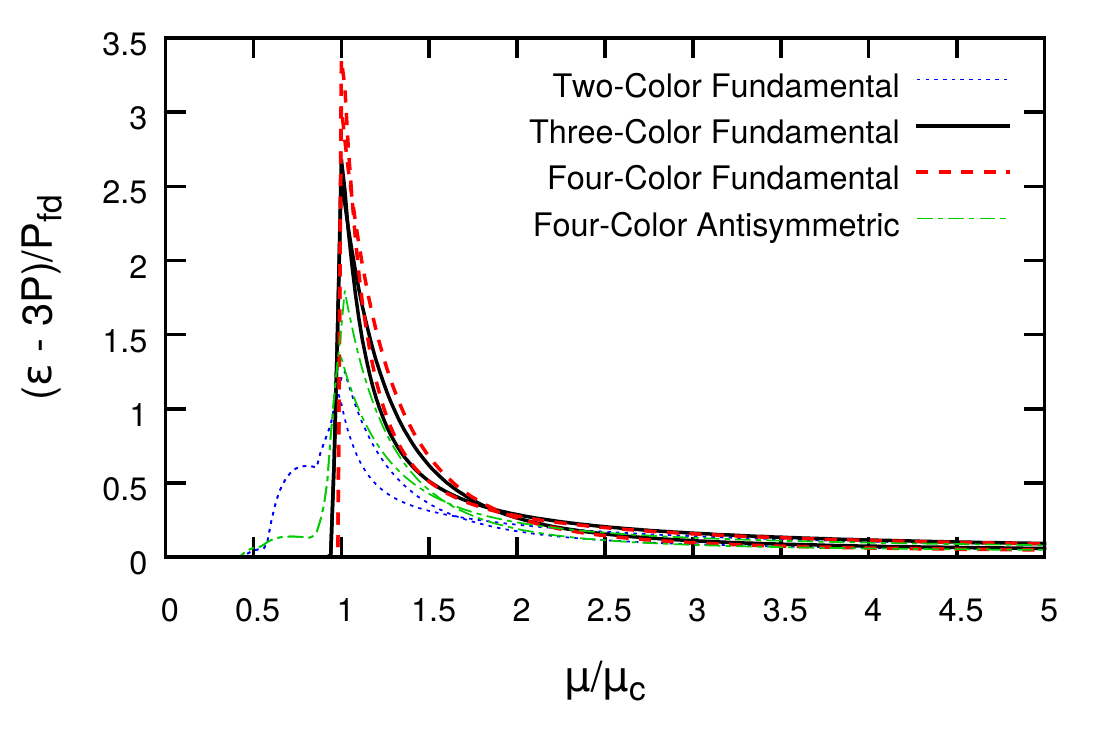}
\caption{Normalized pressure (left) and trace anomaly (right) at $T = 0$ for the 
 two-color, three-color, four-color fundamental, and the four-color antisymmetric theories in \hqcd.  Note that the $\mu$ axis 
 has been scaled by the critical chemical potential (see main text).}
\label{fig: Overlay EoS mu}
\end{figure}

In Figure~\ref{fig: Overlay EoS mu}, 
we show the pressure and trace-anomaly bands at $T = 0$ for all four theories with the $\mu$ axis scaled by the critical 
chemical potential $\mu_{c}$, again, defined to be the average of the matching chemical potential of the upper and lower edge of our 
pQCD band to the HRG equation of state. The value of $\mu_c$ should be regarded as an estimate for the confinement-deconfinement 
transition, whereas the critical chemical potential for the onset transition would be given by the smallest value of 
$m_{i} / r_{i}$, to use the notation of Section~\ref{sec: HRG}.  In the fundamental theories, this value of $m_{i} / r_{i}$ 
corresponds to the lightest baryon mass. Similar to the $\mu=0$ case, the $\mu\neq 0, T=0$ results show similar
trends when scaled appropriately. Again, the different behaviors at low $\mu/\mu_{c}$ are due to the fact that there are 
Nambu--Goldstone bosons composed solely of quarks in the two-color fundamental and four-color antisymmetric theories. Again, 
this was tested by increasing the masses of the lightest particles.

The values of $T_{c} /\sqrt{\sigma}$ and $\mu_{c} / \sqrt{\sigma}$ for our \hqcd\ calculations are given in Table \ref{tab: Tc MuC}. 
While the results suggests that the $T_c$ values for the different theories are within 20 percent of each other, the extracted $\mu_c$ 
values span a much broader range.

We stress that in the four-color antisymmetric 
case with $\lsqrtsigma=0.723$, we were unable to carry out our matching procedure at $\mu = 0$ in the chiral limit.  
We found that in this case, the HRG pressure rose too sharply and never intersected the pQCD pressure-band. Thus, we have 
only plotted the $\lsqrtsigma = 0.527$ results for the four-color antisymmetric theory in our figures.
We feel this is justified for a few reasons. First of all, the values of $\mu_{c}$ are equal within uncertainties
for the two different values of $\lsqrtsigma$. Secondly, in the case where $\lsqrtsigma = 0.723$, we were able to carry out our \hqcd\ 
matching procedure when we increased the mass of the lightest bosons (the pion mass). By varying the pion mass, we were able to
extrapolate to the chiral limit, obtaining a value of $\mu_{c} / \sqrt{\sigma} = 0.3$, which agrees with the value found for 
$\lsqrtsigma = 0.527$. In light of this agreement, and in light of how the four-color antisymmetric theory was the only theory where 
our matching was strained, we conjecture that the true value of $\lsqrtsigma$ in this case is closer to the pure-glue value than it 
is in the real-world, three-color case. We point out that this prediction could be tested in future lattice-gauge-theory calculations.

\begin{figure}[h]
\centering
\includegraphics[width = 0.49\linewidth]{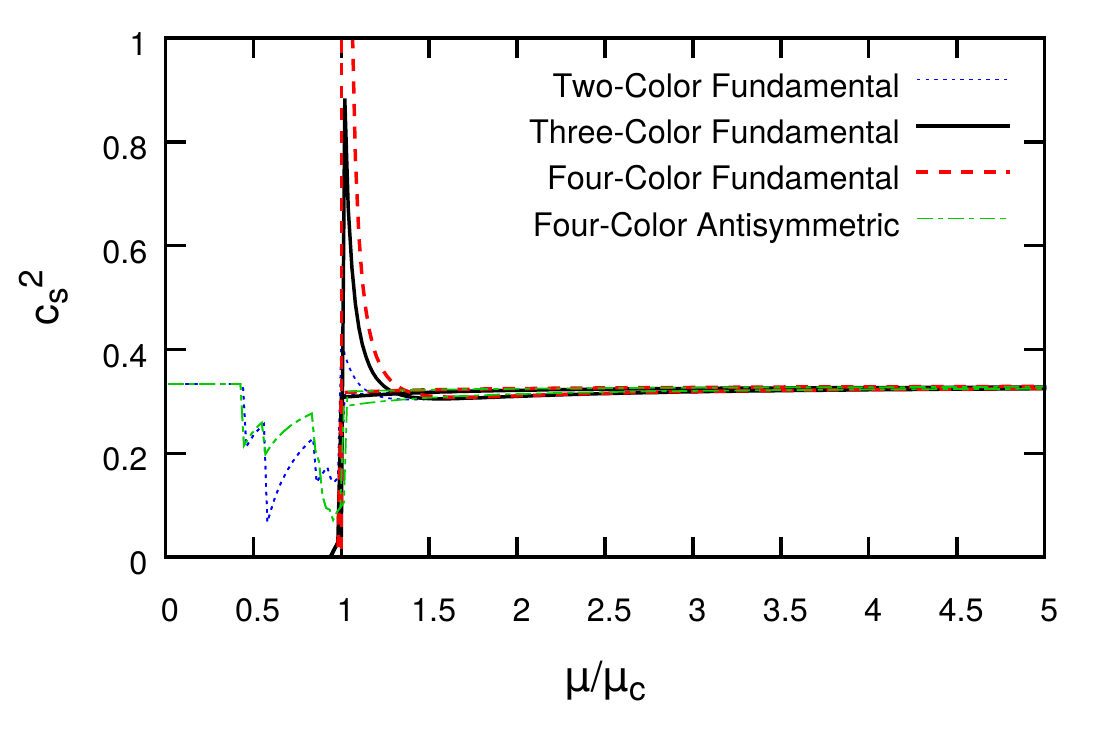}
\caption{The speed of sound squared at $T = 0$ for the 
 two-color, three-color, four-color fundamental, and the four-color antisymmetric theories in \hqcd. Note that the $\mu$ axis 
 has been scaled by the critical chemical potential (see main text). }
\label{fig: Overlay SpdSnd}
\end{figure}

Finally, we have calculated the speed of sound $c_{s}$ at $T = 0$ in all four QCD-like theories using our \hqcd\ scheme, shown 
in Figure~\ref{fig: Overlay SpdSnd}.  We note that in some cases, $c_s$ exceeds the speed of light, and thus these particular 
matching results from \hqcd\ should be considered unphysical (a standard constraint when using cold-nuclear-matter equations of 
state). Nevertheless, our results indicate that it is generally possible to obtain physical equations of state wherein $c_{s}^{2} 
> 1 / 3$ for all fundamental QCD-like theories. This finding could be of interest because restricting $c_{s}^2 < 1 / 3$ has 
previously been noted to be in tension with astrophysical observations \cite{Bedaque:2014sqa}. Again, we point out that this 
is a property which could be tested in future lattice-gauge-theory calculations.
 

%
\begin{table}[h]
\begin{center}
\begin{tabular}{|c|c|c|}
\hline
\em \textbf{Group}, \textbf{Representation}, $\bm{n_{f}}$ & $\,\bm{T_{c} / \sqrt{\sigma}}\,$ & $\,\bm{\mu_{c} / \sqrt{\sigma}}\,$ \\
\hline
SU(2), fundamental, 2 & 0.400 & 3.24 \\
\hline
SU(3), fundamental, 3 & 0.47 & 2.382 \\
\hline
SU(4), fundamental, 2 & 0.44 & 2.853 \\
\hline
SU(4), antisymmetric, 2 ($\Lambda_{\overline{\text{MS}}} / \sqrt{\sigma} = 0.527$) & 0.29 & 5.09 \\
\hline
SU(4), antisymmetric, 2 ($\Lambda_{\overline{\text{MS}}} / \sqrt{\sigma} = 0.723$) & no matching & 5.0 \\
\hline
\end{tabular}
\end{center}
\caption{The ratios $T_{c} / \sqrt{\sigma}$ and $\mu_{c} / \sqrt{\sigma}$ for the theories analyzed in this paper. Errors are 
given by the number of significant figures.}
\label{tab: Tc MuC}
\end{table}
\begin{figure}[h]
\centering
\includegraphics[width = 0.49\linewidth]{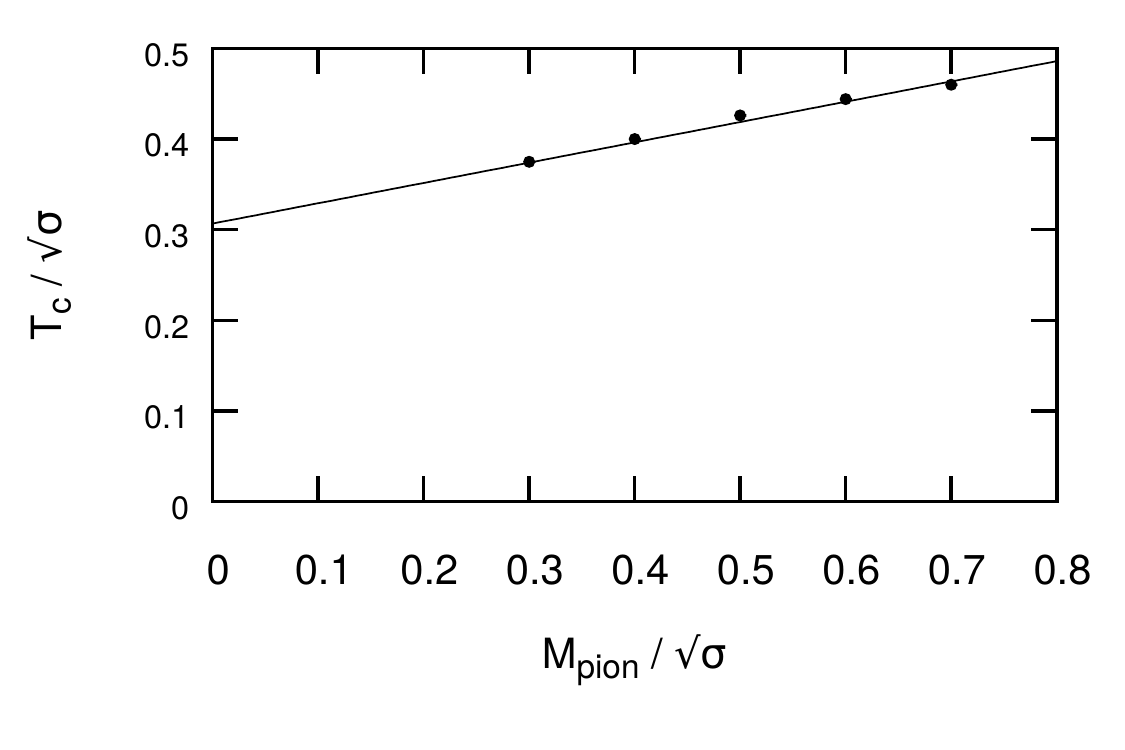}
\caption{Deconfinement-transition temperature $T_{c}$ as a function of pion mass for the four-color, antisymmetric theory in 
 the $\mu = 0$, $\lsqrtsigma = 0.723$ case. The straight line is a fit to the results 
 where matching could be performed, and defines the extrapolation to the chiral limit (see main text).}
\label{fig: MassExtrap}
\end{figure}

\FloatBarrier

\section{Conclusions}

We have calculated the equation of state at non-zero temperatures and densities in a first-principles approach: by matching physics 
from the hadron resonance gas at low energies to perturbative QCD at high energies for two-,\linebreak three-, and four-color `QCD'. 
In particular, our work provides predictions for results in future lattice studies at zero temperature and non-zero chemical potential 
for two-color QCD with two fundamental fermions and four-color QCD with two flavors of fermions in the two-index, antisymmetric 
representation. While some aspects of our study are systematically improvable, we expect the current \hqcd\ results to be sufficiently 
robust that a direct comparison with future lattice-QCD studies in the two- and four-color cases could validate or rule out the \hqcd\ 
method, depending on the quantitative agreement. In the case of agreement, one could thus also reasonably expect \hqcd\ results to be 
quantitatively accurate in the physically-relevant, three-color-QCD case. 

To make our results accessible, we have made them electronically available \cite{Paul:website}. 

\acknowledgements

We thank Gert Aarts,  Tom DeGrand, Simon Hands, Yuzhi Liu, Marco Panero, Andreas Schmitt, and  Aleksi Vuorinen for helpful discussions 
and suggestions.  This work was funded by the Department of Energy, DoE award No.\ DE-SC0008132.

\appendix
\section{Particle Tables}
\label{sec: particle tables}
\begin{table}[h]
\begin{minipage}[b]{0.32\linewidth}\centering
{\large Mesons} \\
\vspace{0.2cm}
\begin{tabular}{|c|c|c|}
\hline
\textbf{Mass/$\sqrt{\sigma}$} & \textbf{Spin} & \textbf{Isospin} \\
\hline
0.00         & 0 & 1 \\ \hline
1.43         & 0 & 0 \\ \hline
1.60         & 1 & 1 \\ \hline
1.86         & 1 & 0 \\ \hline
2.79         & 1 & 0 \\ \hline
3.02         & 2 & 0 \\ \hline
3.06         & 1 & 0 \\ \hline
3.08         & 0 & 0 \\ \hline
3.10         & 1 & 1 \\ \hline
3.14         & 2 & 1 \\ \hline
3.25         & 1 & 1 \\ \hline
3.26         & 0 & 0 \\ \hline
3.38         & 1 & 0 \\ \hline
3.50         & 0 & 1 \\ \hline
3.50         & 0 & 1 \\ \hline
\end{tabular}
\end{minipage}
\hspace{-1.6cm}
\begin{minipage}[b]{0.32\linewidth}\centering
{(continued)} \\
\vspace{0.2cm}
\begin{tabular}{|c|c|c|}
\hline
\textbf{Mass/$\sqrt{\sigma}$} & \textbf{Spin} & \textbf{Isospin} \\
\hline
3.65         & 1 & 1 \\ \hline
3.92         & 2 & 0 \\ \hline
3.93         & 1 & 0 \\ \hline
3.98         & 2 & 1 \\ \hline
3.98         & 3 & 0 \\ \hline
4.02         & 3 & 1 \\ \hline
4.05         & 1 & 1 \\ \hline
4.25         & 0 & 1 \\ \hline
4.25         & 0 & 0 \\ \hline
4.35         & 1 & 1 \\ \hline
4.35         & 1 & 0 \\ \hline
4.86         & 4 & 1 \\ \hline
4.88         & 4 & 0 \\ \hline
5.00         & 1 & 1 \\ \hline
5.00         & 1 & 0 \\ \hline
\end{tabular}
\end{minipage}
\hspace{-0.6cm}
\begin{minipage}[b]{0.32\linewidth}\centering
{\large Baryons} \\
\vspace{0.2cm}
\begin{tabular}{|c|c|c|}
\hline
\textbf{Mass/$\sqrt{\sigma}$} & \textbf{Spin} & \textbf{Isospin} \\
\hline
0.00               & 0 & 0 \\ \hline
1.43               & 0 & 0 \\ \hline
1.86               & 1 & 1 \\ \hline
2.79               & 1 & 1 \\ \hline
3.26               & 0 & 0 \\ \hline
3.06               & 1 & 0 \\ \hline
3.02               & 2 & 0 \\ \hline
3.92               & 2 & 0 \\ \hline
3.93               & 1 & 1 \\ \hline
3.98               & 3 & 1 \\ \hline
4.88               & 4 & 0 \\ \hline
3.08               & 0 & 0 \\ \hline
3.38               & 1 & 1 \\ \hline
5.00               & 1 & 1 \\ \hline
4.25               & 0 & 0 \\ \hline
4.35               & 1 & 0 \\ \hline
\end{tabular}
\end{minipage}
\caption{The included particle spectrum in the two-color fundamental theory.}
\label{tab: two color spectrum}
\end{table}
\begin{table}[h]
\begin{minipage}[b]{0.45\linewidth}\centering
{\large Mesons} \\
\vspace{0.2cm}
\begin{tabular}{|c|c|c|}
\hline
\textbf{Mass/$\sqrt{\sigma}$} & \textbf{Spin} & \textbf{Isospin} \\
\hline
0.00 & 0 & 1 \\ \hline
1.43 & 0 & 0 \\ \hline
1.83 & 1 & 1 \\ \hline
1.86 & 1 & 0 \\ \hline
2.33 & 0 & 1 \\ \hline
2.79 & 1 & 0 \\ \hline
2.94 & 1 & 1 \\ \hline
3.00 & 1 & 1 \\ \hline
3.02 & 2 & 0 \\ \hline
3.06 & 1 & 0 \\ \hline
3.08 & 0 & 0 \\ \hline
3.10 & 0 & 1 \\ \hline
3.14 & 2 & 1 \\ \hline
3.26 & 0 & 0 \\ \hline
3.38 & 1 & 0 \\ \hline
3.45 & 1 & 1 \\ \hline
3.45 & 0 & 1 \\ \hline
3.90 & 1 & 1 \\ \hline
3.92 & 2 & 0 \\ \hline
3.93 & 1 & 0 \\ \hline
3.98 & 2 & 1 \\ \hline
3.98 & 3 & 0 \\ \hline
4.02 & 3 & 1 \\ \hline
4.05 & 1 & 1 \\ \hline
4.25 & 0 & 0 \\ \hline
4.35 & 1 & 0 \\ \hline
4.86 & 4 & 1 \\ \hline
4.88 & 4 & 0 \\ \hline
5.00 & 1 & 1 \\ \hline
5.00 & 1 & 0 \\ \hline
\end{tabular}
\end{minipage}
\hspace{-3cm}
\begin{minipage}[b]{0.45\linewidth}\centering
{\large Baryons} \\
\vspace{0.2cm}
\begin{tabular}{|c|c|c|}
\hline
\textbf{Mass/$\sqrt{\sigma}$} & \textbf{Spin} & \textbf{Isospin} \\
\hline
2.84 & 0 & 1 \\ \hline
3.05 & 1 & 3 \\ \hline
3.47 & 2 & 5 \\ \hline
2.84 & 0 & 5 \\ \hline
3.05 & 1 & 5 \\ \hline
3.47 & 2 & 5 \\ \hline
3.05 & 1 & 3 \\ \hline
3.47 & 2 & 3 \\ \hline
4.10 & 3 & 3 \\ \hline
\end{tabular}
\end{minipage}
\caption{The included particle spectrum in the four-color fundamental theory.}
\label{tab: four color F spectrum}
\end{table}
\begin{table}[h]
\begin{minipage}[b]{0.32\linewidth}\centering
{\large Mesons} \\
\vspace{0.2cm}
\begin{tabular}{|c|c|c|}
\hline
\textbf{Mass/$\sqrt{\sigma}$} & \textbf{Spin} & \textbf{Isospin} \\
\hline
0.00 & 0 & 1 \\ \hline
1.43 & 0 & 0 \\ \hline
1.83 & 1 & 1 \\ \hline
1.86 & 1 & 0 \\ \hline
2.33 & 0 & 1 \\ \hline
2.79 & 1 & 0 \\ \hline
2.94 & 1 & 1 \\ \hline
3.00 & 1 & 1 \\ \hline
3.02 & 2 & 0 \\ \hline
3.06 & 1 & 0 \\ \hline
3.08 & 0 & 0 \\ \hline
3.10 & 0 & 1 \\ \hline
3.14 & 2 & 1 \\ \hline
3.26 & 0 & 0 \\ \hline
3.38 & 1 & 0 \\ \hline
3.45 & 1 & 1 \\ \hline
3.45 & 0 & 1 \\ \hline
3.90 & 1 & 1 \\ \hline
3.92 & 2 & 0 \\ \hline
3.93 & 1 & 0 \\ \hline
3.98 & 2 & 1 \\ \hline
3.98 & 3 & 0 \\ \hline
4.02 & 3 & 1 \\ \hline
4.05 & 1 & 1 \\ \hline
4.25 & 0 & 0 \\ \hline
4.35 & 1 & 0 \\ \hline
4.86 & 4 & 1 \\ \hline
4.88 & 4 & 0 \\ \hline
5.00 & 1 & 1 \\ \hline
5.00 & 1 & 0 \\ \hline
\end{tabular}
\end{minipage}
\hspace{-0.8cm}
\begin{minipage}[b]{0.32\linewidth}\centering
{\large Diquarks} \\
\vspace{0.2cm}
\begin{tabular}{|c|c|c|}
\hline
\textbf{Mass/$\sqrt{\sigma}$} & \textbf{Spin} & \textbf{Isospin} \\
\hline
0.00 & 0 & 1 \\ \hline
1.43 & 0 & 1 \\ \hline
1.86 & 1 & 0 \\ \hline
2.79 & 1 & 0 \\ \hline
3.02 & 2 & 1 \\ \hline
3.06 & 1 & 1 \\ \hline
3.08 & 0 & 1 \\ \hline
3.26 & 0 & 1 \\ \hline
3.38 & 1 & 0 \\ \hline
3.92 & 2 & 1 \\ \hline
3.93 & 1 & 0 \\ \hline
3.98 & 3 & 0 \\ \hline
4.25 & 0 & 1 \\ \hline
4.35 & 1 & 1 \\ \hline
4.88 & 4 & 1 \\ \hline
5.00 & 1 & 0 \\ \hline
\end{tabular}
\end{minipage}
\hspace{-0.8cm}
\begin{minipage}[b]{0.32\linewidth}\centering
{\large Baryons} \\
\vspace{0.2cm}
\begin{tabular}{|c|c|c|}
\hline
\textbf{Mass/$\sqrt{\sigma}$} & \textbf{Spin} & \textbf{Isospin} \\
\hline
0.71 & 0 & 0 \\ \hline
1.55 & 1 & 1 \\ \hline
3.24 & 2 & 2 \\ \hline
5.77 & 3 & 3 \\ \hline
\end{tabular}
\end{minipage}
\caption{The mesons, diquarks, and baryons in the four-color antisymmetric theory. (See Table \ref{tab: four color AS2
 tetraquark spectrum} for remaining particles in this theory.)}
\label{tab: four color AS2 spectrum}
\end{table}
\begin{table}
\begin{minipage}[b]{0.245\linewidth}\centering
{\large Tetraquarks, di-mesons, and diquark-mesons} \\
\vspace{0.2cm}
\begin{tabular}{|c|c|c|}
\hline
\textbf{Mass/$\sqrt{\sigma}$} & $\bm{g_{\text{\textbf{Spin}}}}$ & $\bm{g_{\text{\textbf{Isospin}}}}$ \\
\hline
0.00 & 1  & 1  \\ \hline
1.43 & 1  & 1  \\ \hline
1.86 & 3  & 3  \\ \hline
2.79 & 3  & 3  \\ \hline
2.86 & 1  & 1  \\ \hline
3.02 & 5  & 5  \\ \hline
3.06 & 3  & 3  \\ \hline
3.08 & 1  & 1  \\ \hline
3.26 & 1  & 1  \\ \hline
3.29 & 3  & 3  \\ \hline
3.38 & 3  & 3  \\ \hline
3.72 & 9  & 9  \\ \hline
3.92 & 5  & 5  \\ \hline
3.93 & 3  & 3  \\ \hline
3.98 & 7  & 7  \\ \hline
4.21 & 3  & 3  \\ \hline
4.25 & 1  & 1  \\ \hline
4.35 & 3  & 3  \\ \hline
4.45 & 5  & 5  \\ \hline
4.49 & 3  & 3  \\ \hline
4.51 & 1  & 1  \\ \hline
4.65 & 9  & 9  \\ \hline
4.69 & 1  & 1  \\ \hline
4.81 & 3  & 3  \\ \hline
4.88 & 9  & 9  \\ \hline
4.89 & 15 & 15 \\ \hline
4.92 & 9  & 9  \\ \hline
4.95 & 3  & 3  \\ \hline
5.00 & 3  & 3  \\ \hline
5.12 & 3  & 3  \\ \hline
5.24 & 9  & 9  \\ \hline
5.35 & 5  & 5  \\ \hline
5.36 & 3  & 3  \\ \hline
5.40 & 7  & 7  \\ \hline
\end{tabular}
\end{minipage}
\begin{minipage}[b]{0.245\linewidth}\centering
{(continued)} \\
\vspace{0.2cm}
\begin{tabular}{|c|c|c|}
\hline
\textbf{Mass/$\sqrt{\sigma}$} & $\bm{g_{\text{\textbf{Spin}}}}$ & $\bm{g_{\text{\textbf{Isospin}}}}$ \\
\hline
5.57 & 9  & 9  \\ \hline
5.68 & 1  & 1  \\ \hline
5.78 & 18 & 18 \\ \hline
5.79 & 9  & 9  \\ \hline
5.81 & 15 & 15 \\ \hline
5.84 & 21 & 21 \\ \hline
5.85 & 9  & 9  \\ \hline
5.87 & 3  & 3  \\ \hline
6.05 & 28 & 28 \\ \hline
6.08 & 15 & 15 \\ \hline
6.11 & 5  & 5  \\ \hline
6.11 & 3  & 3  \\ \hline
6.12 & 9  & 9  \\ \hline
6.14 & 3  & 3  \\ \hline
6.17 & 10 & 10 \\ \hline
6.21 & 9  & 9  \\ \hline
6.29 & 5  & 5  \\ \hline
6.31 & 9  & 9  \\ \hline
6.32 & 3  & 3  \\ \hline
6.35 & 1  & 1  \\ \hline
6.40 & 15 & 15 \\ \hline
6.43 & 3  & 3  \\ \hline
6.44 & 9  & 9  \\ \hline
6.46 & 3  & 3  \\ \hline
6.52 & 1  & 1  \\ \hline
6.64 & 3  & 3  \\ \hline
6.70 & 15 & 15 \\ \hline
6.71 & 9  & 9  \\ \hline
6.74 & 27 & 27 \\ \hline
6.76 & 30 & 30 \\ \hline
6.86 & 9  & 9  \\ \hline
6.94 & 25 & 25 \\ \hline
6.95 & 15 & 15 \\ \hline
6.98 & 15 & 15 \\ \hline
\end{tabular}
\end{minipage}
\begin{minipage}[b]{0.245\linewidth}\centering
{(continued)} \\
\vspace{0.2cm}
\begin{tabular}{|c|c|c|}
\hline
\textbf{Mass/$\sqrt{\sigma}$} & $\bm{g_{\text{\textbf{Spin}}}}$ & $\bm{g_{\text{\textbf{Isospin}}}}$ \\
\hline
6.99 & 9  & 9  \\ \hline
7.00 & 40 & 40 \\ \hline
7.01 & 3  & 3  \\ \hline
7.04 & 24 & 24 \\ \hline
7.06 & 7  & 7  \\ \hline
7.14 & 9  & 9  \\ \hline
7.18 & 5  & 5  \\ \hline
7.19 & 3  & 3  \\ \hline
7.24 & 7  & 7  \\ \hline
7.27 & 5  & 5  \\ \hline
7.30 & 15 & 15 \\ \hline
7.31 & 12 & 12 \\ \hline
7.33 & 1  & 1  \\ \hline
7.36 & 21 & 21 \\ \hline
7.37 & 15 & 15 \\ \hline
7.41 & 9  & 9  \\ \hline
7.43 & 3  & 3  \\ \hline
7.51 & 1  & 1  \\ \hline
7.61 & 3  & 3  \\ \hline
7.63 & 3  & 3  \\ \hline
7.67 & 27 & 27 \\ \hline
7.73 & 9  & 9  \\ \hline
7.79 & 9  & 9  \\ \hline
7.83 & 25 & 25 \\ \hline
7.85 & 15 & 15 \\ \hline
7.86 & 9  & 9  \\ \hline
7.89 & 35 & 35 \\ \hline
7.90 & 66 & 66 \\ \hline
7.94 & 27 & 27 \\ \hline
7.95 & 49 & 49 \\ \hline
7.96 & 9  & 9  \\ \hline
8.02 & 15 & 15 \\ \hline
8.06 & 9  & 9  \\ \hline
8.08 & 3  & 3  \\ \hline
\end{tabular}
\end{minipage}
\begin{minipage}[b]{0.245\linewidth}\centering
{(continued)} \\
\vspace{0.2cm}
\begin{tabular}{|c|c|c|}
\hline
\textbf{Mass/$\sqrt{\sigma}$} & $\bm{g_{\text{\textbf{Spin}}}}$ & $\bm{g_{\text{\textbf{Isospin}}}}$ \\
\hline
8.14 & 9  & 9  \\ \hline
8.17 & 5  & 5  \\ \hline
8.18 & 3  & 3  \\ \hline
8.23 & 7  & 7  \\ \hline
8.26 & 30 & 30 \\ \hline
8.27 & 15 & 15 \\ \hline
8.28 & 9  & 9  \\ \hline
8.33 & 21 & 21 \\ \hline
8.38 & 9  & 9  \\ \hline
8.50 & 1  & 1  \\ \hline
8.60 & 3  & 3  \\ \hline
8.70 & 9  & 9  \\ \hline
8.80 & 45 & 45 \\ \hline
8.81 & 27 & 27 \\ \hline
8.86 & 63 & 63 \\ \hline
8.92 & 15 & 15 \\ \hline
8.93 & 9  & 9  \\ \hline
8.98 & 21 & 21 \\ \hline
9.13 & 9  & 9  \\ \hline
9.23 & 27 & 27 \\ \hline
9.25 & 3  & 3  \\ \hline
9.35 & 9  & 9  \\ \hline
9.76 & 81 & 81 \\ \hline
9.88 & 27 & 27 \\ \hline
\end{tabular}
\end{minipage}
\caption{The included tetraquarks, di-mesons, and diquark-mesons in the four-color antisymmetric theory. (There is one of each 
 of these particle types for each line in this table.) Here, $g_{\text{Spin}}$ and $g_{\text{Isospin}}$ are the total spin and 
 isospin degeneracies, respectively.  As noted above in Section \ref{sec: HRG}, we need not determine how all of the 
 four-quark-object degrees of freedom break up into spin and isospin multiplets because of the mass degeneracy.}
\label{tab: four color AS2 tetraquark spectrum}
\end{table}
\FloatBarrier
\bibliographystyle{unsrt}
\bibliography{References}

\begin{thebibliography}{10}

\bibitem{Borsanyi:2010cj}
{Borsanyi, Szabolcs and Endrodi, Gergely and Fodor, Zoltan and Jakovac, Antal
  and Katz, Sandor D. and others}.
\newblock {The QCD equation of state with dynamical quarks}.
\newblock {\em JHEP}, 1011:077, 2010.

\bibitem{Borsanyi:2013bia}
Szabocls Borsanyi, Zoltan Fodor, Christian Hoelbling, Sandor~D. Katz, Stefan
  Krieg, et~al.
\newblock {Full result for the QCD equation of state with 2+1 flavors}.
\newblock {\em Phys.Lett.}, B730:99--104, 2014.

\bibitem{Bazavov:2014pvz}
A.~Bazavov and others (HotQCD~Collaboration).
\newblock {Equation of state in (2+1)-flavor QCD}.
\newblock {\em Phys.Rev.}, D90(9):094503, 2014.

\bibitem{Cristoforetti:2012su}
Marco Cristoforetti, Francesco Di~Renzo, and Luigi Scorzato.
\newblock {New approach to the sign problem in quantum field theories: High
  density QCD on a Lefschetz thimble}.
\newblock {\em Phys.Rev.}, D86:074506, 2012.

\bibitem{Aarts:2013uxa}
Gert Aarts, Lorenzo Bongiovanni, Erhard Seiler, Denes Sexty, and Ion-Olimpiu
  Stamatescu.
\newblock {Controlling complex Langevin dynamics at finite density}.
\newblock {\em Eur.Phys.J.}, A49:89, 2013.

\bibitem{Sexty:2013ica}
Dénes Sexty.
\newblock {Simulating full QCD at nonzero density using the complex Langevin
  equation}.
\newblock {\em Phys.Lett.}, B729:108--111, 2014.

\bibitem{Aarts:2014bwa}
Gert Aarts, Erhard Seiler, Denes Sexty, and Ion-Olimpiu Stamatescu.
\newblock {Simulating QCD at nonzero baryon density to all orders in the
  hopping parameter expansion}.
\newblock 2014.

\bibitem{Aarts:2014fsa}
Gert Aarts, Felipe Attanasio, Benjamin Jäger, Erhard Seiler, Denes Sexty,
  et~al.
\newblock {QCD at nonzero chemical potential: recent progress on the lattice}.
\newblock 2014.

\bibitem{deForcrand:2014tha}
Philippe de~Forcrand, Jens Langelage, Owe Philipsen, and Wolfgang Unger.
\newblock {The lattice QCD phase diagram in and away from the strong coupling
  limit}.
\newblock {\em Phys.Rev.Lett.}, 113:152002, 2014.

\bibitem{Laine:2006cp}
Mikko Laine and York Schroder.
\newblock {Quark mass thresholds in QCD thermodynamics}.
\newblock {\em Phys.Rev.}, D73:085009, 2006.

\bibitem{Huovinen:2009yb}
Pasi Huovinen and Pter Petreczky.
\newblock {QCD Equation of State and Hadron Resonance Gas}.
\newblock {\em Nucl.Phys.}, A837:26--53, 2010.

\bibitem{Kurkela:2009gj}
Aleksi Kurkela, Paul Romatschke, and Aleksi Vuorinen.
\newblock {Cold Quark Matter}.
\newblock {\em Phys.Rev.}, D81:105021, 2010.

\bibitem{Langelage:2008dj}
Jens Langelage, Gernot Munster, and Owe Philipsen.
\newblock {Strong coupling expansion for finite temperature Yang-Mills theory
  in the confined phase}.
\newblock {\em JHEP}, 0807:036, 2008.

\bibitem{Vuorinen:2003fs}
A.~Vuorinen.
\newblock {The Pressure of QCD at finite temperatures and chemical potentials}.
\newblock {\em Phys.Rev.}, D68:054017, 2003.

\bibitem{Kajantie:2002wa}
K.~Kajantie, M.~Laine, K.~Rummukainen, and Y.~Schroder.
\newblock {The Pressure of hot QCD up to g6 ln(1/g)}.
\newblock {\em Phys.Rev.}, D67:105008, 2003.

\bibitem{Kajantie:2003ax}
K.~Kajantie, M.~Laine, K.~Rummukainen, and Y.~Schroder.
\newblock {Four loop vacuum energy density of the SU(N(c)) + adjoint Higgs
  theory}.
\newblock {\em JHEP}, 0304:036, 2003.

\bibitem{Blaizot:2003iq}
J.P. Blaizot, E.~Iancu, and A.~Rebhan.
\newblock {On the apparent convergence of perturbative QCD at high
  temperature}.
\newblock {\em Phys.Rev.}, D68:025011, 2003.

\bibitem{Haque:2014rua}
Najmul Haque, Aritra Bandyopadhyay, Jens~O. Andersen, Munshi~G. Mustafa,
  Michael Strickland, et~al.
\newblock {Three-loop HTLpt thermodynamics at finite temperature and chemical
  potential}.
\newblock {\em JHEP}, 1405:027, 2014.

\bibitem{Braaten:1995jr}
Eric Braaten and Agustin Nieto.
\newblock {Free energy of QCD at high temperature}.
\newblock {\em Phys.Rev.}, D53:3421--3437, 1996.

\bibitem{vanRitbergen:1997va}
T.~van Ritbergen, J.A.M. Vermaseren, and S.A. Larin.
\newblock {The Four loop beta function in quantum chromodynamics}.
\newblock {\em Phys.Lett.}, B400:379--384, 1997.

\bibitem{Hands:2006ve}
Simon Hands, Seyong Kim, and Jon-Ivar Skullerud.
\newblock {Deconfinement in dense 2-color QCD}.
\newblock {\em Eur.Phys.J.}, C48:193, 2006.

\bibitem{Kogut:2000ek}
J.B. Kogut, Misha~A. Stephanov, D.~Toublan, J.J.M. Verbaarschot, and
  A.~Zhitnitsky.
\newblock {QCD - like theories at finite baryon density}.
\newblock {\em Nucl.Phys.}, B582:477--513, 2000.

\bibitem{Kapusta:2006pm}
J.I. Kapusta and Charles Gale.
\newblock {\em {Finite-temperature field theory: Principles and applications}}.
\newblock 2006.

\bibitem{Agashe:2014kda}
K.A. Olive et~al.
\newblock {Review of Particle Physics}.
\newblock {\em Chin.Phys.}, C38:090001, 2014.

\bibitem{Meyer:2004jc}
Harvey~B. Meyer and Michael~J. Teper.
\newblock {Glueball Regge trajectories and the pomeron: A Lattice study}.
\newblock {\em Phys.Lett.}, B605:344--354, 2005.

\bibitem{Lucini:2008vi}
Biagio Lucini and Gregory Moraitis.
\newblock {The Running of the coupling in SU(N) pure gauge theories}.
\newblock {\em Phys.Lett.}, B668:226--232, 2008.

\bibitem{Fritzsch:2012wq}
Patrick Fritzsch, Francesco Knechtli, Bjorn Leder, Marina Marinkovic, Stefan
  Schaefer, et~al.
\newblock {The strange quark mass and Lambda parameter of two flavor QCD}.
\newblock {\em Nucl.Phys.}, B865:397--429, 2012.

\bibitem{Bali:2013kia}
Gunnar~S. Bali, Francis Bursa, Luca Castagnini, Sara Collins, Luigi Del~Debbio,
  et~al.
\newblock {Mesons in large-N QCD}.
\newblock {\em JHEP}, 1306:071, 2013.

\bibitem{Hands:2007uc}
Simon Hands, Peter Sitch, and Jon-Ivar Skullerud.
\newblock {Hadron Spectrum in a Two-Colour Baryon-Rich Medium}.
\newblock {\em Phys.Lett.}, B662:405--412, 2008.

\bibitem{Lewis:2011zb}
Randy Lewis, Claudio Pica, and Francesco Sannino.
\newblock {Light Asymmetric Dark Matter on the Lattice: SU(2) Technicolor with
  Two Fundamental Flavors}.
\newblock {\em Phys.Rev.}, D85:014504, 2012.

\bibitem{Adkins:1983ya}
Gregory~S. Adkins, Chiara~R. Nappi, and Edward Witten.
\newblock {Static Properties of Nucleons in the Skyrme Model}.
\newblock {\em Nucl.Phys.}, B228:552, 1983.

\bibitem{Jenkins:1993zu}
Elizabeth~Ellen Jenkins.
\newblock {Baryon hyperfine mass splittings in large N QCD}.
\newblock {\em Phys.Lett.}, B315:441--446, 1993.

\bibitem{DeGrand:2012hd}
Thomas DeGrand.
\newblock {Lattice baryons in the 1/N expansion}.
\newblock {\em Phys.Rev.}, D86:034508, 2012.

\bibitem{Appelquist:2014jch}
T.~Appelquist et~al.
\newblock {Composite bosonic baryon dark matter on the lattice: SU(4) baryon
  spectrum and the effective Higgs interaction}.
\newblock {\em Phys.Rev.}, D89:094508, 2014.

\bibitem{DeGrand:2014U}
Thomas DeGrand, Yuzhi Liu, Ethan Neil, Yigal Shamir, and Benjamin Svetitsky.
\newblock {Spectroscopy of SU(4) gauge theory with two flavors of sextet
  fermions}.
\newblock {To appear}.

\bibitem{DeGrand:2014cea}
Thomas DeGrand, Yuzhi Liu, Ethan~T. Neil, Yigal Shamir, and Benjamin Svetitsky.
\newblock {Spectroscopy of SU(4) lattice gauge theory with fermions in the two
  index anti-symmetric representation}.
\newblock 2014.

\bibitem{Bolognesi:2006ws}
Stefano Bolognesi.
\newblock {Baryons and Skyrmions in QCD with Quarks in Higher Representations}.
\newblock {\em Phys.Rev.}, D75:065030, 2007.

\bibitem{Peskin:1980gc}
Michael~E. Peskin.
\newblock {The Alignment of the Vacuum in Theories of Technicolor}.
\newblock {\em Nucl.Phys.}, B175:197--233, 1980.

\bibitem{Kurkela:2014vha}
{Kurkela, Aleksi and Fraga, Eduardo S. and Schaffner-Bielich, J\"{u}rgen and
  Vuorinen, Aleksi}.
\newblock {Constraining neutron star matter with Quantum Chromodynamics}.
\newblock {\em Astrophys.J.}, 789:127, 2014.

\bibitem{Kojo:2014rca}
Toru Kojo, Philip~D. Powell, Yifan Song, and Gordon Baym.
\newblock {Phenomenological QCD equation of state for massive neutron stars}.
\newblock 2014.

\bibitem{Panero:2009tv}
Marco Panero.
\newblock {Thermodynamics of the QCD plasma and the large-N limit}.
\newblock {\em Phys.Rev.Lett.}, 103:232001, 2009.

\bibitem{Bedaque:2014sqa}
Paulo~F. Bedaque and Andrew~W. Steiner.
\newblock {Sound velocity bound and neutron stars}.
\newblock 2014.

\bibitem{Paul:website}
The results may be obtained from the web page of one of the authors,
  \url{http://hep.itp.tuwien.ac.at/~paulrom/}.

\end{thebibliography}

\end{document}